\documentclass[preprint,12pt]{elsarticle}

\usepackage{verbatim} 
\usepackage{epsfig}

\usepackage{amssymb} 

\usepackage{mathtools}
\usepackage{lineno}

\usepackage{tikz}        
\usepackage{etoolbox}

\usepackage[caption=false]{subfig}
\usepackage[utf8]{inputenc}
\usepackage{hyperref}
\usepackage{url}
\biboptions{sort&compress}
\begin{document}

\newcommand*{\hwplotB}{\raisebox{3pt}{\tikz{\draw[red,dashed,line 
width=3.2pt](0,0) -- 
(5mm,0);}}}

\newrobustcmd*{\mydiamond}[1]{\tikz{\filldraw[black,fill=#1] (0,0) -- 
(0.1cm,0.2cm) --
(0.2cm,0) -- (0.1cm,-0.2cm);}}

\newrobustcmd*{\mytriangleleft}[1]{\tikz{\filldraw[black,fill=#1] (0,0.15cm) -- 
(-0.3cm,0) -- (0,-0.15cm);}}
\definecolor{Blue}{cmyk}{1.,1.,0,0} 

\begin{frontmatter}



\title{Microscopic dynamics at the Running of the Bulls (San Fermín Festival) 
in the context of the Social Force Model}

\author[label1]{F.E.~Cornes}
\author[label3]{G.A.~Frank}
\author[label1,label2]{C.O.~Dorso}
\address[label1]{Departamento de F\'\i sica, Facultad de Ciencias 
Exactas y Naturales, Universidad de Buenos Aires, Pabell\'on I, Ciudad 
Universitaria, 1428 Buenos Aires, Argentina.}
\address[label2]{Instituto de F\'\i sica de Buenos Aires, Pabell\'on I, Ciudad 
Universitaria, 1428 Buenos Aires, Argentina.}
\address[label3]{ Unidad de Investigaci\'on y Desarrollo de las 
Ingenier\'\i as, Universidad Tecnol\'ogica Nacional, Facultad Regional Buenos 
Aires, Av. Medrano 951, 1179 Buenos Aires, Argentina.}



\begin{abstract}

This research explores the dynamics of the emergency evacuation during the 
``Running of the Bulls'' festival (Spain, 2013). As people run to escape from 
danger, many pedestrians stumble and fall down, while others will try to 
pass over them. We carefully examined three specific recordings of the 
running, that show this kind of behavior. We developed a microscopic model 
mimicking the stumbling mechanism in the context of the Social Force Model 
(SFM). In our model, ``moving'' individuals can suddenly switch to a ``fallen'' 
state when they are in the vicinity of a fallen individual. We arrived to the 
conclusion that the presence of a fallen pedestrian increases dramatically the 
falling probability of the pedestrians nearby. Also, the product between the 
local density gradient and the velocity of each pedestrian appears as a relevant 
indicator for an imminent fall. We call this the pedestrian 
``falling susceptibility ($f_s$)''. \\
\end{abstract}

\begin{keyword}

Pedestrian dynamics \sep Social Force Model \sep 
Stumbling mechanism


\PACS 45.70.Vn \sep 89.65.Lm


\end{keyword}

\end{frontmatter}



\section{\label{motivation}Introduction}

The matter of emergency evacuation is of obvious importance in common life. In 
the last decades, a growing interest appeared in the research of the pedestrian 
dynamics during an emergency evacuation. The proper understanding of the 
evacuation dynamics will allow safer facilities when designing common spaces.\\

The evacuation through narrow pathways or doorways appears as one of the most 
studied scenarios in the literature. Many experiments and numerical 
simulations have been carried out for a better understanding of the crowd 
behavior in a variety of situations. These investigations focus on the door 
width as the main cause of the overcrowding during the evacuation process 
\cite{Haghani_2019,Sarvi_2019, Boltes-Seyfried_2018,Zuriguel_2016,
Seyfried-Boltes_2014,Hoogendoorn_2012, Seyfried-Boltes_2011,Boltes_2009,
Seyfried-Schadschneider_2009}. Therefore, it is pointed as the key magnitude 
affecting the evacuation performance. \\

The dynamic of emergency evacuations through narrow doors is dominated by 
``blockings''. Research has shown that a small group of pedestrians close to the 
door can be responsible for blocking the way to the rest of 
the crowd by forming an arch-like metastable structure, or 
\textit{blocking cluster}, yielding long lasting delays 
\cite{Dorso1,Dorso2,Dorso6,Dorso7}. 
Also, the pressure inside the bulk is considered to be of decisive importance in 
the breaking process of the blocking clusters \cite{Helbing1, Dorso1, Dorso6, 
Dorso7, Zuriguel_2013, Daniels_2011, Zuriguel_2013_b, Daniels_2016_a, 
Daniels_2016_b, Daniels_2018}. \\

The pedestrian's anxiety \cite{Dorso7}, the presence of obstacles \cite{Dorso3} 
and the possibility of fallings \cite{Dorso5} are all relevant issues that 
affect the formation of blocking clusters in the context of narrow doors 
(bottlenecks). However, blocking clusters are expected to become irrelevant out 
of this context \cite{Dorso1}, and therefore, the connection between the 
aforementioned issues and the evacuation performance appears somehow obscure. \\

The emergency evacuations in the absence of blocking clusters are, from our 
point of view, an unattended matter throughout the literature. Thus, we decided 
to place the attention on those situations where the blocking clusters are quite 
improbable (\textit{i.e.} wide openings), but fallings are 
still a relevant issue. Specifically, our 
investigation focuses on three real-life situations (empirical approach) of 
the ``Running of the Bulls'' event (2013). The videos capture the instances 
where participants stumble and fall during the rush in different wide door 
scenarios \cite{precorridor,corridor,postcorridor}. \\

Our first objective is to acquire reliable data from real stumbling events, 
and use this data to build a formalism to recognize situations 
in which tripping becomes highly probable within the context of the 
SFM. We assume that pedestrians behave as either ``moving'' or ``fallen'' 
individuals. The transition probabilities from the ``moving'' behavior to the 
``fallen'' one corresponds to those from the video recordings. No transition 
from ``fallen'' to ``moving'' is permitted under this approach. \\

We emphasize the critical attitude that we sustained along the investigation. 
Our SFM model was tested before proceeding to other hypothetical scenarios. 
Results showed fairly good agreement with the analyzed videos. \\

The investigation is organized as follows. A brief review of the basic ``Social 
Force Model'' can be found in Section~\ref{background}. In 
Section~\ref{sec:experimental} we present the real-life situations considered in 
our investigation, while in Section~\ref{sec:num_simul} we provide details on 
the numerical simulations. Section~\ref{sec:results} present the results from 
both the empirical and numerical analyses. The corresponding conclusions are 
summarized in Section~\ref{sec:conclusions}.

\section{\label{background}Background}

\subsection{\label{sfm}The social force model}

This investigation handles the pedestrians dynamics in the context of the 
``Social Force Model'' (SFM) \cite{Helbing1}. The SFM exploits the idea 
that human motion depends on the people's own desire to reach a certain 
destination at a given velocity, as well as other environmental factors 
\cite{Helbing4}. The former is modeled by a force called the ``desire force'', 
while the latter is represented by ``social forces'' and ``granular forces''. 
These forces enter the equation of motion as follows

\begin{equation}
m_i\,\displaystyle\frac{d\mathbf{v}^{(i)}}{dt}=\mathbf{f}_d^{(i)}
+\displaystyle\sum_{j=1}^{N}\displaystyle\mathbf{f}_s^{(ij)}
+\displaystyle\sum_ {
j=1}^{N}\mathbf{f}_g^{(ij)}\label{eq_mov}
\end{equation}

\noindent where the $i,j$ subscripts correspond to any two pedestrians in the 
crowd. $\mathbf{v}^{(i)}(t)$ means the current velocity of the pedestrian  
$(i)$, while $\mathbf{f}_d$ and $\mathbf{f}_s$ correspond to the ``desired 
force'' and the ``social force'', respectively. $\mathbf{f}_g$ is the friction 
or granular force. \\

The $\mathbf{f}_d$ describes the pedestrians own desire to reach a specific 
target position at the desired velocity $v_d$. But, due to environmental 
factors 
(\textit{i.e.} obstacles, visibility), he (she) actually moves at the current 
velocity $\mathbf{v}^{(i)}(t)$. Thus, he (she) will accelerate (or decelerate) 
to reach any desired velocity $v_d$ that will make him (her) feel more 
comfortable. Thus, in the Social Force Model, the desired force reads 
\cite{Helbing1}

\begin{equation}
        \mathbf{f}_d^ {(i)}(t) =  
m_i\,\displaystyle\frac{v_d^{(i)}\,\mathbf{e}_d^
{(i)}(t)-\mathbf{v}^{(i)}(t)}{\tau} \label{desired}
\end{equation}

\noindent where $m_i$ is the mass of the pedestrian $i$ and $\tau$ represents 
the relaxation time needed to reach the desired velocity. $\mathbf{e}_d$ 
is the unit vector pointing to the target position. Detailed values for $m_i$ 
and $\tau$ can be found in 
Refs.~\cite{Helbing1,Dorso3,Sticco_2020,Sticco_2021}.\\

The ``social force'' $\mathbf{f}_s(t)$ represents the socio-psychological 
tendency of the pedestrians to preserve their \emph{private sphere}. The 
spatial preservation means that a repulsive feeling exists between two 
neighboring pedestrians, or, between the pedestrian and the walls 
\cite{Helbing1,Helbing4}. This repulsive feeling becomes stronger as people get 
closer to each other (or to the walls). Thus, in the context of the Social 
Force 
Model, this tendency is expressed as 

\begin{equation}
        \mathbf{f}_s^{(ij)} = A_i\,e^{(r_{ij}-d_{ij})/B_i}\mathbf{n}_{ij} 
        \label{social}
\end{equation}

\noindent where $(ij)$ corresponds to any two pedestrians, or to the 
pedestrian-wall interaction. $A_i$ and $B_i$ are two fixed parameters (see 
Ref.~\cite{Dorso1}). The distance $r_{ij}=r_i+r_j$ is the sum of the 
pedestrians radius, while $d_{ij}$ is the distance between the center of mass 
of the pedestrians $i$ and $j$. $\mathbf{n}_{ij}$ means the unit vector in the 
$\vec{ji}$ direction. For the case of repulsive feelings with the walls, 
$d_{ij}$ corresponds to the shortest distance between the pedestrian and the 
wall, while $r_{ij}=r_i$ (see Refs.~\cite{Helbing1,Helbing4}).  \\

It is worth mentioning that the Eq.~(\ref{social}) is also valid if two 
pedestrians are in contact (\textit{i.e.} $r_{ij}>d_{ij}$), but its meaning is 
somehow different. In this case, $\mathbf{f}_s$ represents a body repulsion, as 
explained in Ref.~\cite{Dorso5}.\\

The granular force $\mathbf{f}_g$ included in Eq.~(\ref{eq_mov}) corresponds 
to the sliding friction between pedestrians in contact, or, between pedestrians 
in contact with the walls. The expression for this force is 

\begin{equation}
        \mathbf{f}_g^{(ij)} = 
\kappa\,(r_{ij}-d_{ij})\,\Theta(r_{ij}-d_{ij})\,\Delta
\mathbf{v}^{(ij)}\cdot\mathbf{t}_{ij} 
        \label{granular}
\end{equation}

\noindent where $\kappa$ is a fixed parameter. The function 
$\Theta(r_{ij}-d_{ij})$ is zero when its argument is negative (that is, 
$r_{ij}<d_{ij}$) and equals unity for any other case (Heaviside function). 
$\Delta\mathbf{v}^{(ij)}\cdot\mathbf{t}_{ij}$ represents the difference between 
the tangential velocities of the sliding bodies (or between the individual and 
the walls). Detailed values for all the parameters ($m_i$, $\tau$, $A_i$, 
$B_i$, $\kappa$) can be found in Refs.~\cite{Helbing1,Sticco_2020,Sticco_2021}.  
\\

\subsection{\label{human}Clusters}

Many previous works showed that clusters of pedestrians play a fundamental 
role during the evacuation process \cite{Dorso1,Dorso2,Dorso7}. In 
this sense, researchers demonstrated that there exist a relation between 
them and the clogging up of people. Clusters of pedestrians can be defined as 
the set of individuals that for any member of the group (say, $i$) there exists 
at least another member belonging to the same group ($j$) in contact with the 
former. Thus, we define a \textit{human cluster} ($C_g$) following the 
mathematical formula given in Ref.~\cite{strachan}. 

\begin{equation}
C_g:p_i~\epsilon~ C_g \Leftrightarrow \exists~ p_j~\epsilon~C_g / r_{ij} < 
(R_i+R_j) \label{ec-cluster}
\end{equation}

\noindent where ($p_i$) indicate the \textit{ith} pedestrian and $R_i$ is his 
(her) radius (shoulder width). This means that $C_g$ is a set of pedestrians 
that interact not only with the social force, but also with physical forces 
(\textit{i.e.} friction force).\\

\section{\label{sec:experimental}Experimental data}


This investigation focuses on real life videos of human rush incidents. We 
fully analyzed three specific episodes of the 
``Running of the Bulls'' festival in San Fermín (Spain), as occurred in 2013. 
The videos are freely available in YouTube 
\cite{precorridor,corridor,postcorridor} and show a crowd of pedestrians running 
away from bulls. \\

\begin{center}
\begin{figure}
\centering \includegraphics[scale = 0.3]{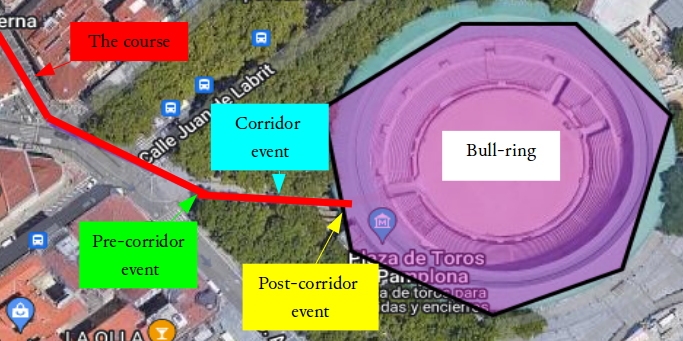} \\
\begin{tabular}{|@{\hspace{2mm}}c|@{\hspace{2mm}}c|@{\hspace{2mm}}c|}
\multicolumn{3}{c}{Pre-corridor scenario}\\
\hline
\includegraphics[scale = 0.32]{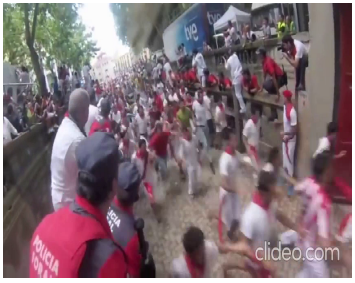} & 
\includegraphics[scale = 0.32]{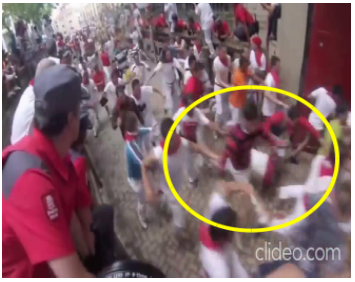} & 
\includegraphics[scale = 0.32]{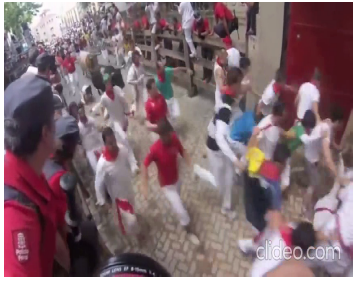}
  \\
a (t=0\,s) & b (t=1.5\,s) & c (t=2.5\,s) \\\hline
  \multicolumn{3}{c}{Corridor scenario}\\
  \hline
\includegraphics[scale = 0.32]{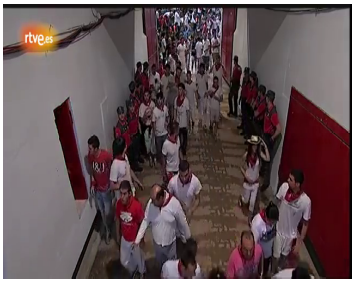} & 
\includegraphics[scale = 0.32]{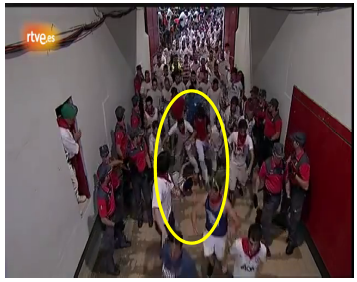} & 
\includegraphics[scale = 0.32]{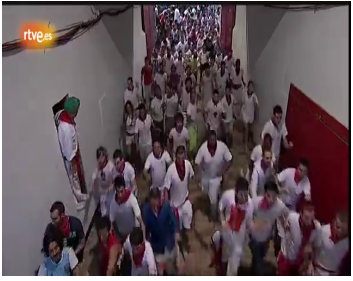}\\
d (t=5\,s) & e (t=20\,s) & f (t=30\,s) \\
\hline
 \multicolumn{3}{c}{Post-Corridor scenario}\\
  \hline
\includegraphics[scale = 0.32]{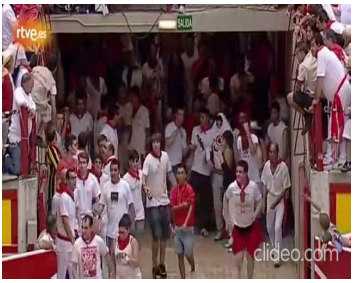} & 
\includegraphics[scale = 0.32]{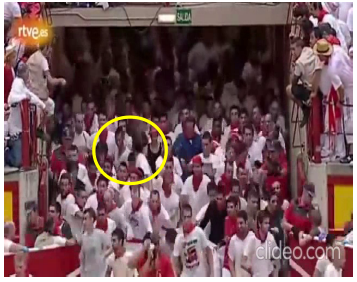} & 
\includegraphics[scale = 0.32]{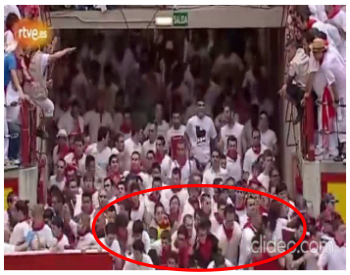}\\
g (t=10\,s) & h (t=25\,s) & i (t=30\,s) \\
\hline
\end{tabular}
\caption{(Color on-line only) (Upper) Layout of the street and locations where 
the incidents took place. Pre-corridor event (first row), corridor event 
(second row) and post-corridor event (third row). The snapshots address the time 
line of the event. The corresponding time-stamp is indicated below
the snapshots. The yellow and red circles indicate the region where the first 
pedestrian stumbles and where the stumbling pedestrians block the exit, 
respectively. In the last picture (lower right) many pedestrians pile-up (out 
of the scene).}
\label{tab:snapshots_real_video}
\end{figure}
\end{center}

The bull-run takes place along the streets of the city 
of Pamplona (Spain). Participants escape from the bulls and some of 
them stumble and fall during the rush. Fig.~\ref{tab:snapshots_real_video} 
captures three instances of the running when, at least, one pedestrian 
stumbles and falls. Each row in Fig.~\ref{tab:snapshots_real_video} corresponds 
to a falling episode. The snapshots on each row capture the time sequence of 
the falling event (see captions for details). \\

For convenience, we will refer to each episode in 
Fig.~\ref{tab:snapshots_real_video} as the ``pre-corridor event'' 
\cite{precorridor}, the ``corridor event'' \cite{corridor} and the 
``post-corridor event'' \cite{postcorridor}. The latter occurs at the 
entrance of the bull-ring, but still in the corridor limits. 

The following are the feature actions taking place in the events:

\begin{itemize}
 \item Pre-corridor: One pedestrian stumbles and falls during the rush. Few 
seconds after, he (she) manages to get up and continues his (her) way out. 
Neighboring pedestrians run without major incidents.
 \item Corridor: Three pedestrians stumble and fall during the event. Notice 
that, unlike the pre-corridor event, they stumble in a higher density 
scenario. Fortunately, they get up quickly and continue their escaping route. 
 \item Post-corridor: A cluster of more than forty pedestrians stumble and fall 
during the rush. The video shows that the majority of falls are correlated in 
both space and time. Unlike the other two events, the first fallen pedestrian is 
unable to stand up quickly, triggering a massive stumble. Consequently, the 
crowd evolved towards a completely block scenario (see red circle in 
Fig.~\ref{tab:snapshots_real_video}). 
\end{itemize}

The very first inspection of the events brings out the fact 
that the local density around the first falling pedestrian is a crucial feature 
to the forthgoing events. That is, the higher the local density, the greater 
tendency towards a massive stumble. We will leave the above matter to 
Section~\ref{sec:local_den} and for now we will focus on the data acquisition 
procedure (see Section~\ref{sec:data_acquisition}). We will further introduce 
some definitions in Section~\ref{sec:micro_obs}.

\subsection{\label{sec:data_acquisition}Data acquisition}

In order to quantify the three incidents exhibited in 
Fig.~\ref{tab:snapshots_real_video}, we sampled the movies at a rate of 8 
frames per second (attaining a time interval between successive images of 
0.125 seconds). The criteria for choosing this sampling period was: 1) The 
period should be large enough to include (at least) one fallen individual and to 
achieve a reasonable amount of measurements, 2) during this period, the camera 
must be fixed to perform the right perspective correction. \\

The total number of acquired images for the pre-corridor event was 32, while 
the other two events attained 240 images each. This allowed a semi-manual 
tracking of the pedestrians by means of the ImageJ software \cite{Rasband}. We 
made sure that any pedestrian trajectory included at least 10 data points. 

\subsection{\label{sec:micro_obs}Definitions of relevant magnitudes}

The following are the most meaningful magnitudes in the context of our 
investigation.  

\begin{figure*}[!ht]
\centering
\hspace{5mm}
\subfloat[]{\includegraphics[
width=0.4\columnwidth]{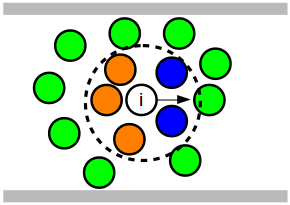}
}
\hspace{5mm}
\subfloat[]{\includegraphics[
width=0.4\columnwidth]{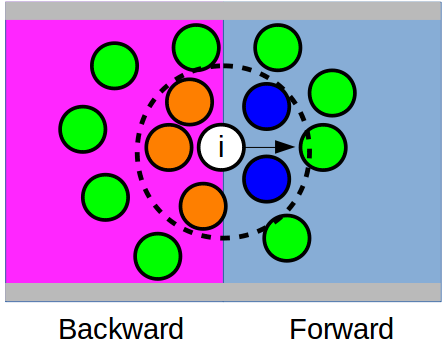}
}
\caption{\label{fig:diagramas_medicion} (Color on-line only) Schematic diagram 
for the individuals in the corridor. The circles represent pedestrians moving 
from left to right. The upper and lower lines represent the walls of the 
corridor. The dashed circle corresponds to the measurement area around 
pedestrian $i$. In this case, he (she) is surrounded by five individuals 
($N_i=5$). Therefore, the local density is $\rho=5/\pi$~(p/m$^2)$  
(see Eq.~\ref{eqn_rho}). Also, notice that three of these neighbors are located 
backward with respect to pedestrian $i$ (orange circles), while the other 
two are placed forward (blue circles). Therefore, the local gradient density 
around pedestrian $i$ is $\nabla\rho_i=1$ (see Eq.~\ref{eqn_grad}). } 
\end{figure*}

\begin{itemize}
 \item The local density at any position $\vec{r} = (x,y)$ and time $t$ 
was computed as follows

\begin{equation}
 \rho_i(\vec{r},t) = \frac{N_i(\vec{r},t)}{\pi R^2}
\label{eqn_rho}
\end{equation}

\noindent where $N_i(\vec{r},t)$ represents the number of neighbors of 
pedestrian $i$ within a fix radius $R=1\,$m (see 
Fig.~\ref{fig:diagramas_medicion}). 

 \item The instantaneous speed (modulus) for each runner was computed as the 
symmetric incremental ratio 

 \begin{equation}
v_i(t) = \frac{|\vec{r_i}(t+\Delta t)-\vec{r_i}(t-\Delta t)|}{2\Delta t}
\label{eqn_vel}
\end{equation}

\noindent for $\Delta t=0.125\,$s, 

 \item The local density gradient around a pedestrian $i$ (modulus) was defined 
as follows

\begin{equation}
\nabla\rho_i(t)=|N_f(t)-N_b(t)|
\label{eqn_grad}
\end{equation}

\noindent where $N_f$ and $N_b$ represent the number of moving pedestrians 
located in the forward and backward directions, respectively 
(see Fig.~\ref{fig:diagramas_medicion} for details). We only considered moving 
pedestrians for the gradient computation. That is, fallen individuals were 
excluded from the computation. The $N_f$ and $N_b$ counts only considered 
pedestrians within the distance $R=1\,$m to pedestrian $i$ (see 
Fig.~\ref{fig:diagramas_medicion}). 
 \item We defined the product
 
 \begin{equation}
f_{s,i} = v_i\nabla\rho_i
\label{eqn_prod}
\end{equation}
 
 \noindent as a ``susceptibility'' to the fall. This ``susceptibility'' 
accounts for the compound nature of the falling: the pedestrian's own speed 
($v_i$) and the pushing environment ($\nabla\rho_i$).\\

\end{itemize}

The sampling area was located in the middle of the pathway, as 
depicted in Fig.~\ref{fig:Medicion_dens_local}. As already mentioned, each 
magnitude was recorded every $0.125\,$s. 

\section{\label{sec:num_simul}Numerical simulations}

\subsection{\label{sec:boundary}Boundary and initial conditions}

In order to study this process, we recreated the ``Running 
of the bulls'' event (see Section~\ref{sec:experimental}) by placing from 100 up 
to 700 pedestrians inside a straight corridor of width $w=5\,$m (similar to the 
width at the entrance of the bull-ring) and length $L=50\,$m. The global density 
ranged from $\rho=0.5$ to 3 p/m$^2$. This is similar to the local density 
values measured in the analyzed scenes of the San Fermín festival (see 
Fig.~\ref{fig:local_density}).  \\

The individuals were initially distributed at random positions along the 
simulation box, with random initial velocities according to a Gaussian 
distribution with null mean value. If certain conditions were met (see 
Section~\ref{sec:types}), any moving pedestrian could switch his (her) behavior 
to the ``fallen'' pedestrian behavior. Fallen pedestrians were those that 
remained at a fix position until the end of the simulation process. \\

\begin{figure}
\centering
\includegraphics[scale=0.4]{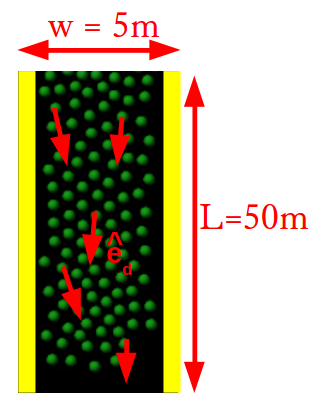}
\caption{\label{fig:diag_sim}(Color on-line only) Snapshot of simulated 
evacuation process through a $5\, $m\,$\times\,50\,$m straight corridor. 
$\mathbf{\hat{e}_d}$ points directly to the exit (say, the entrance of the 
bull-ring at the bottom of this scheme). All the individuals correspond to 
``moving pedestrians''. The left and right yellow lines represent the walls of 
the corridor.}
\end{figure}

The desired velocity $v_d$ was the same for all the individuals, meaning 
that all of them had the same anxiety level. 
The desired direction $\mathbf{\hat{e}_d}$ was updated at each time step in 
order to point to the exit (say, the entrance of the bull-ring) as depicted in 
Fig.~\ref{fig:diag_sim}. 

\subsection{\label{sec:types}Moving and fallen pedestrians}

We assumed two behavioral patterns, as already mentioned: ``moving'' individuals 
and ``fallen'' individuals. The former are those that move according to 
Eq.~(\ref{eq_mov}). The latter are those that are not able to 
move at all until the end of the evacuation process. Moving pedestrians, 
however, are able to switch to the fallen behavior, but fallen pedestrians 
always remain in the same condition. \\

The falling transition was implemented as follows. First, we computed 
the ``fallen susceptibility'' (say, $f_s^{(i)}=v^{(i)}\nabla\rho^{(i)}$) for 
each moving pedestrian, at instance $i$ (see Eq.~\ref{eqn_prod}). Secondly, we 
switched the state of each ``moving'' individual to the ``fallen'' category with 
the following probabilities

\begin{equation}
p(f_s^{(i)}) = \left\{\begin{matrix}
p_{F}(f_s^{(i)}) & \mathrm{if~there~ is~ a~ fallen~ individual~ nearby}\\ 
p_{NF}(f_s^{(i)}) & \mathrm{if~ there~ is~ not~ a~ fallen~ individual~ nearby} 
\end{matrix}\right. \label{eq:prob_sim} 
\end{equation}

\noindent where fallen neighbors are considered to be within a radius of 
1\,m. That is, we considered those fallen neighbors located 
backward and forward the pedestrian $i$. Notice that $p_F$ and 
$p_{NF}$ depend on the susceptibility $f_s^{(i)}$, and are obtained from the 
fitting of experimental data (see Section.~\ref{sec:after} for details on the 
computation of these probabilities). The time-step between instances was 0.5~s. 
\\

The moving pedestrians were allowed to pass through the fallen ones. In order to 
make this possible, we assumed that no social repulsion was present between the 
moving and fallen individuals. See Ref.~\cite{Dorso5} for details.

\subsection{Simulation software}
\label{sec:lammps}

The simulations were implemented on the {\sc Lammps} molecular dynamics 
simulator \cite{plimpton}. {\sc Lammps} was set to run on multiple processors. 
The chosen time integration scheme was the velocity Verlet algorithm with a 
time step of $10^{-4}\,$s. Any other parameter was the same as in previous 
works (see Refs.~\cite{Dorso3,Dorso4}). \\

We performed between 30 and 100 simulations for each scenario in order to get 
enough data for statistical analyses (see figures caption for details). Data 
was sampled at time intervals of 0.05~s along 100~s. The recorded magnitudes 
were the pedestrian's positions and velocities for each evacuation 
process. 

\section{\label{sec:results}Results}

We present in this section the main results of our investigation. The section 
is 
divided into two parts. In the first part (Section~\ref{sec:emp_meas}), we 
present the empirical results extracted from each incident. In the second 
part (Section~\ref{sec:num_sim}), we compare the empirical results with 
numerical simulations. 

\subsection{\label{sec:emp_meas}Empirical results}

The analysis of the videos is divided into two separate Sections. The first one 
examines the crowd behavior previous to the first fallen pedestrian (within 
the sight of the film), while the second one focuses on the dynamics following 
the fall. These are two naturally different stages, since the former 
corresponds to an homogeneous crowds, while the latter deals with two behavioral 
patterns (\textit{i.e.} with ``moving'' and ``fallen'' pedestrians).

\subsubsection{\label{sec:local_den}Before the first fall}

\begin{figure*}[!ht]
\hspace{5mm}
\subfloat[Measuring diagram\label{fig:Medicion_dens_local}]{\includegraphics[
width=0.35\columnwidth]{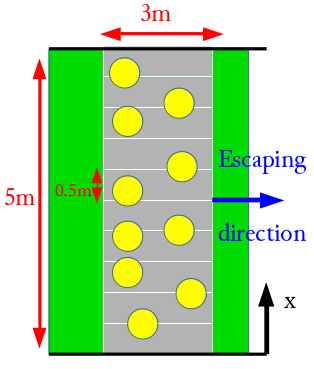}
}
\hspace{-1mm}
\subfloat[Pre-corridor\label{fig:densidad_pre_corridor}]{\includegraphics[
width=0.45\columnwidth]{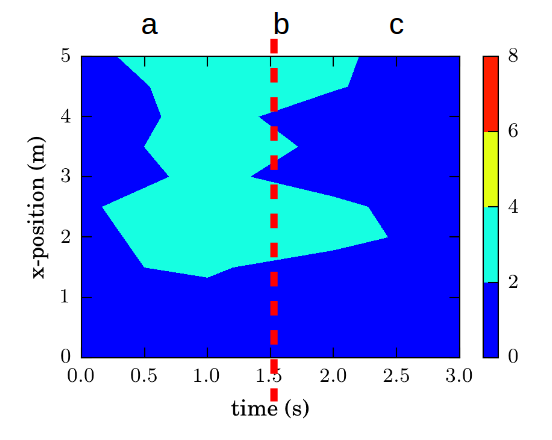}
}
\hspace{-1mm}
\subfloat[Corridor\label{fig:densidad_pasillo}]{
\includegraphics[width=0.46\columnwidth]{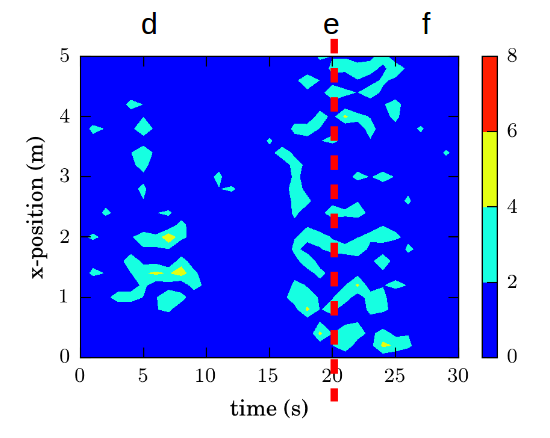}
}
\hspace{-1mm}
\subfloat[Post-Corridor\label{fig:densidad_entrada}]{
\includegraphics[width=0.45\columnwidth]{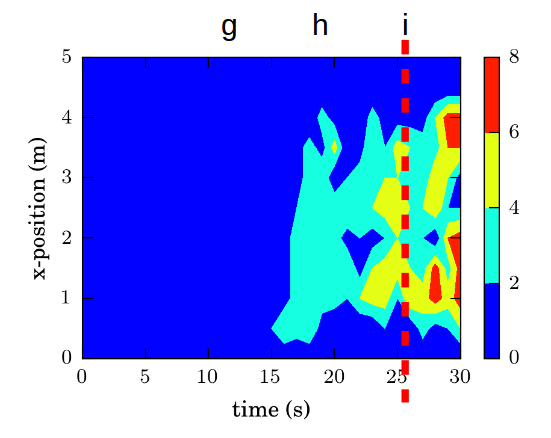}
}
\caption{\label{fig:local_density}(Color on-line only) (a) Schematic diagram 
for individuals during the running. In order to minimize boundary effects, the 
measurements were taken from the middle of each scenario (gray box). This box 
was binned in cells of size $3\,\mathrm{m}\times0.5\,\mathrm{m}$. (b-c-d) 
Density contour lines along time for each scenario. The maps were splined to get 
smoother curves. Each cell of size $0.5\,\mathrm{s}\times0.5\,\mathrm{m}$ 
corresponds to the occupancy density at time $t$ in position $x$ (see the 
measuring diagram). The scale bar on the right is expressed in 
pedestrians/m$^2$. The vertical red dashed line represents the time of the first 
fall. The letters above each plot mean the corresponding snapshots in 
Fig.~\ref{tab:snapshots_real_video}. The horizontal and vertical axis represent 
the time and the pedestrian's $x-$location in each corridors, respectively. The 
walls are located at $x=0$ and $x=5$ (bottom and top of each figure).} 
\end{figure*}

The analysis started by slicing the scenes shown in 
Fig.~\ref{tab:snapshots_real_video}. We made 
the corresponding perspective corrections and computed the magnitudes defined 
previously in Section~\ref{sec:micro_obs}. Fig.~\ref{fig:Medicion_dens_local} 
schemes the procedure for the computation of each observable. \\


Figs.~\ref{fig:densidad_pre_corridor}-\ref{fig:densidad_entrada} show the local 
density ($\rho$) pattern along time for each scenario (see caption for 
details). We can notice a common behavior from the contour maps in 
Fig.~\ref{fig:local_density}: the first fall occurs \emph{after} an increase 
in the local density. Or, in other words, an increase in the density somehow 
\textit{anticipates} a forthcoming fall. This does not mean that \emph{any} 
increment in the density naturally yields a falling incident, but sets an 
environmental condition for expecting it (in the context of the San Fermín 
festival). \\

A closer inspection of the scenes \emph{before} the falls shows that the 
natural 
tendency of the pedestrians is to preserve some distance between them while 
running, presumably because of trying to dodge other runners. However, the 
upcoming bulls disturb this tendency towards a more chaotic scenario. 
Consequently, there is no time left for trying to dodge others, and the density 
increases. \\

In order to quantify the above observation, we measured the pedestrians' 
velocity ($v_i$) and the local density gradient ($\nabla\rho$), as defined in 
Section~\ref{sec:micro_obs}. Both magnitudes and some comments on its time 
evolution can be found in \ref{appendix_1}. However, we actually prefer to 
focus on the susceptibility $f_s$ as defined in Section~\ref{sec:micro_obs}. 
This is, from our point of view, a more meaningful magnitude to express the 
disturbed scenario. \\

\begin{center}
\begin{figure*}[!ht]
\hspace{0mm}
\subfloat[Pre-corridor\label{fig:mapa_producto_pre_pasillo}]{\includegraphics[
width=0.35\columnwidth]{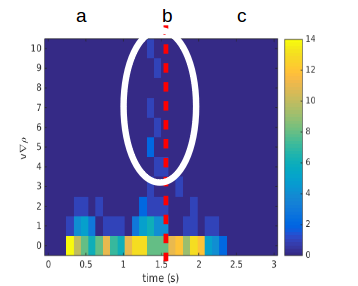}
}
\hspace{-8mm}
\subfloat[Corridor\label{fig:mapa_producto_pasillo}]{\includegraphics[
width=0.35\columnwidth]{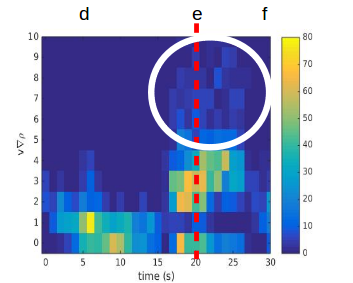}
}
\hspace{-8mm}
\subfloat[Post-Corridor\label{fig:mapa_producto_entrada}]{
\includegraphics[width=0.35\columnwidth]{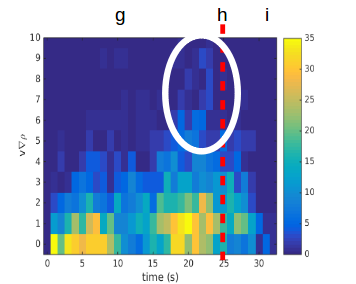}
}
\caption{\label{tab:observables} (Color on-line only) 2-D histograms for the 
falling susceptibility ($v\nabla\rho$) along time for each scenario. The scale 
bar on the right corresponds to the number of pedestrians exposed to 
$v\nabla\rho$ ($y$-axis) at time $t$ ($x$-axis). The bin size 
was $0.125\,\mathrm{s}\times1\,\mathrm{people~m/s}$. The vertical red 
dashed line represents the time-stamp of the first fall. The horizontal and 
vertical axis represent the time and the falling susceptibility, respectively. 
For clarity reasons, the upper limit of each color bar is different. The 
letters above each plot mean the corresponding snapshots in 
Fig.~\ref{tab:snapshots_real_video}.} 
\end{figure*}
\end{center}

Fig.~\ref{tab:observables} shows the susceptibility histograms as a function of 
time for each scenario (see caption for details). Likewise the local density 
patterns, we can notice a common behavior from the histograms: the first fall 
occurs \emph{after} an increase of the number of pedestrians exposed to high 
susceptibility values (see white circles in Fig.~\ref{tab:observables}). \\

Notice that the white circles in Fig.~\ref{tab:observables} do not enclose a 
large amount of events (\textit{i.e.} many counts) but those that are 
significant to the falling process. This means that only a few pedestrians 
attain the susceptibility level to (probably) produce a falling incident.\\

The susceptibility appears as a better indicator for an imminent fall with 
respect to the local density. Fig.~\ref{tab:observables} provides a rough 
threshold (say, $f_s\approx 4$) pointing to an imminent fall. The increase in 
the local density stands for a likely falling condition, although we can not 
set 
any clear threshold for the San Fermín context. \\

\subsubsection{\label{sec:after}After the first fall}

\begin{figure}
\centering
\includegraphics[scale=0.5]{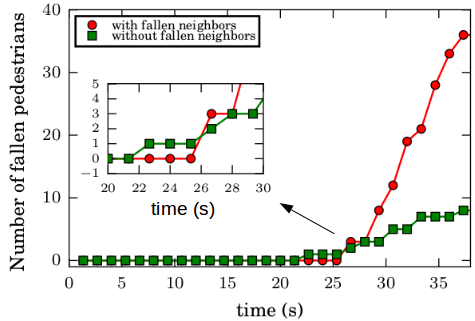}
\caption{\label{fig:plot_con_sin_caidos_alrededor} (Color on-line only) 
Empirical number of fallen pedestrians with (red circles) and without (green 
squares) fallen neighbors as a function of time in the post-corridor scene.}
\end{figure}

We will now focus on the ``fallen'' pedestrians. We first computed the number 
of stumbles in presence (or not) of fallen neighbors as a function of time. 
This is shown in Fig.~\ref{fig:plot_con_sin_caidos_alrededor} for the 
post-corridor scenario only. The pre-corridor and corridor scenes are not shown 
because the scarcity of data did not allow a reliable analysis. \\

As can be seen in Fig.~\ref{fig:plot_con_sin_caidos_alrededor}, near 40 
pedestrians stumble and fall in the vicinity of, at least, one fallen neighbor 
at the end of the process. But this number barely reaches 10 individuals in the 
absence of a neighboring fallen pedestrian. Thus, it becomes clear that the 
presence of at least one fallen neighbor increases dramatically the ratio of 
casualties along time. \\

\begin{figure}
\centering
\includegraphics[scale=0.5]{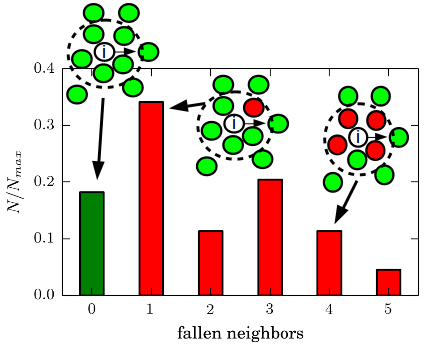}
\caption{\label{fig:plot_caidos_vs_caidos_alrededor} (Color on-line only)  
Normalized cumulative number of fallen pedestrians as a function of the number 
of fallen neighbors in the post-corridor scenario. The count of fallen 
neighbors was done within a fix radius $R=1\,$m. The diagrams above the bins 
are schematic representations for each case. The dashed circle corresponds to 
the measurement area around pedestrian $i$. The plot is normalized with respect 
to the total number of fallen pedestrians. The mean number of fallen 
individuals in this sample distribution is $\mu=1.83$. The bin error is 
expected to be $\sigma=\sqrt{p\,(1-p)/6}$ for $p\approx\mu/6$. }
\end{figure}

We further examined the number of fallen neighbors around any new falling 
pedestrian. The sample probabilities are exhibited in 
Fig.~\ref{fig:plot_caidos_vs_caidos_alrededor}. We estimate a sampling error of 
$\sigma\approx 0.2$ for each bin (see caption for details). This means that the 
case of only one fallen neighbor is actually the significant one. More than one 
fallen neighbors may occur, but its relevancy is still in question according to 
our data.  \\

The fact that fallen neighbors increase dramatically the possibility of new 
fallings turned our investigation to the following working hypothesis: the 
presence of ``fallen neighbors'' increases the chance of falling, in addition 
to the expectation for susceptible (moving) individuals alone. This means that 
those  environment attaining ``fallen individuals'' should be studied 
separately from the ``non-fallen'' ones. \\  

We first computed separately the falling probabilities for 
individuals that belong to an environment with, at least, one fallen neighbor, 
and with no fallen neighbors at all. We studied both situations as a function 
of the susceptibility $f_s$. The following is the expression used for computing 
both falling probabilities ($p_{F}$ and $p_{NF}$, see Eq.~\ref{eq:prob_sim}): 

\begin{equation}
 p(f_s) = \frac{\displaystyle \sum_{t=0}^{T} 
N_F(f_s)}{\displaystyle\sum_{t=0}^{T}N_T(f_s)}
 \label{eqn_prob}\\
\end{equation}

\noindent where $N_F$ and $N_T$ represent the number of falling pedestrians and 
the total number of pedestrians (either fallen and moving ones) attaining a 
falling susceptibility $f_s$, respectively. Both sums run over all the discrete 
time samples (see Section~\ref{sec:data_acquisition} for details). Notice that 
the value of $f_s=v\nabla\rho$ was always sampled just before the fall occurs. 
\\

\begin{figure}
\centering
\includegraphics[scale=0.9]{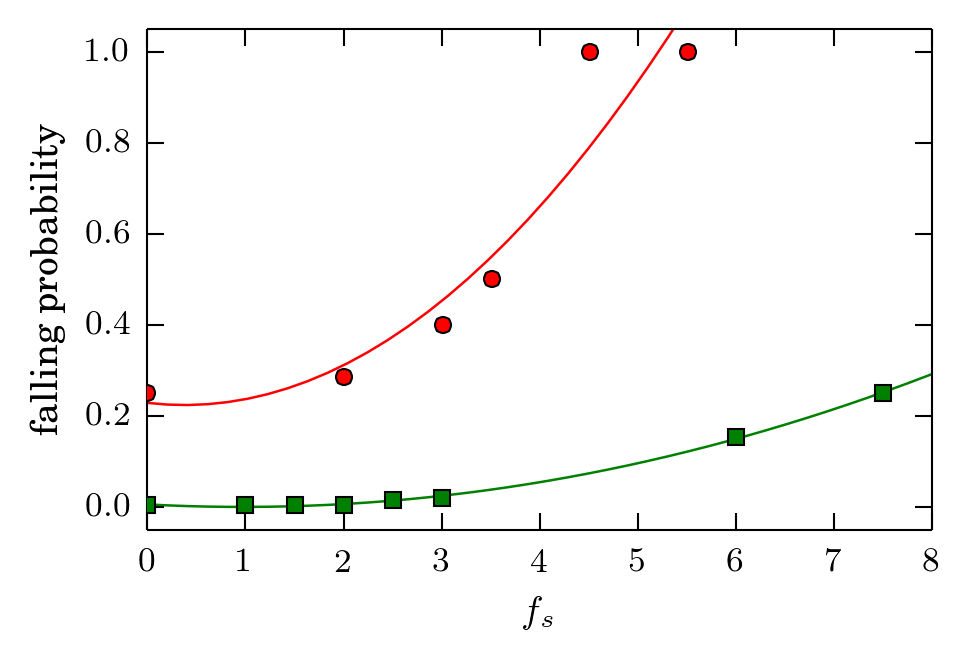}
\caption{\label{fig:plot_proba} (Color on-line only) Falling probability as a 
function of the pedestrian's susceptibility ($f_s=v\nabla\rho$) with (red 
circles) and without (green squares) neighboring fallen pedestrians (see 
Eq.~\ref{eqn_prob}). Solid lines correspond to the quadratic fitting of each 
data set: $p_{F}(s)=0.33\,f_s^2-0.025\,f_s+0.229$ (red curve) and 
$p_{NF}(s)=0.006\,f_s^2-0.011\,f_s+0.001$ 
(green curve). }
\end{figure}

Fig.~\ref{fig:plot_proba} shows the falling probability as a function of the 
susceptibility ($f_s$) in the post-corridor scene (see caption for 
details). It can be seen that the falling probability fits accurately a 
quadratic function with respect to $f_s$ in both environmental conditions (say, 
with and without fallen neighbors). Moreover, Fig.~\ref{fig:plot_proba} is in 
agreement with the rough threshold ($f_s\approx 4$) observed in 
Fig.~\ref{tab:observables} for an imminent falling. We can now confirm that the 
probability of falling increases rapidly beyond this value. \\ 

Interestingly, both curves exhibit a similar qualitative shape regardless of 
the environmental condition. The presence of fallen neighbors, however, 
introduces, 
at least, an additional $20\%$ of chances for the new fall to occur. 
This additional probability due to a fallen neighbor increases as $f_s$ 
becomes larger, as can be noticed from the comparison of the slopes in 
Fig.~\ref{fig:plot_proba}.

\subsubsection*{\label{sec:rem_exp}Concluding remarks from the empirical 
results}

Our major conclusions from this Section are as follows. First, two different 
causes of stumbles can be noticed during the rush: a few individuals fall as 
a consequence of their interaction between other moving pedestrians 
only. However, the majority stumble in presence of, at least, another 
fallen pedestrian. \\

As a second conclusion, we noticed that the falling probability increases as 
the susceptibility $f_s=v\nabla\rho$ (say, the product between the local 
density 
gradient and the velocity of each pedestrian) increases. Thus, fallings are more 
likely to occur whenever people rush away from danger (\emph{i.e} the bulls) or 
embody a striking situation (\emph{i.e} high density gradient). These are the 
commonly observed situations in the San Ferm\'\i n festival.

\subsection{\label{sec:num_sim}Numerical results}

In this Section we widen the empirical picture of the San Fermín festival by 
means of numerical simulations. Our first aim is to mimic this situation within 
the context of the SFM (see Section~\ref{background} for details). We introduce 
falling probabilities to the basic SFM, according to estimates from 
Section~\ref{sec:emp_meas}. We then investigate possible scenarios according to 
the SFM. 

\subsubsection{The morphology in the bulk of the crowd}
\label{sec:bulk}

\begin{center}
\begin{figure}[!ht]
\begin{tabular}{|@{\hspace{0mm}}c|@{\hspace{0mm}}c|}
\multicolumn{2}{c}{$v_d=3\,$m/s}\\
\hline
\includegraphics[scale = 0.52]{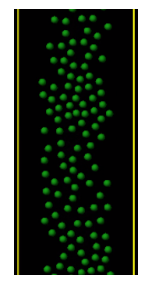} & 
\includegraphics[scale = 0.52]{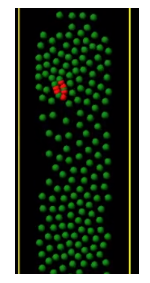} \\
$\rho~=~1\,$p/m$^2$ & $\rho~=~2\,$p/m$^2$ \\
\hline
\end{tabular}
\hspace{10mm}
\begin{tabular}{|@{\hspace{0mm}}c|@{\hspace{0mm}}c|}
\multicolumn{2}{c}{$v_d=5\,$m/s}\\
\hline
\includegraphics[scale = 0.52]{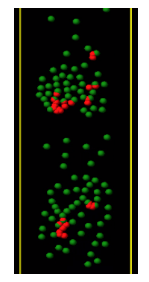} & 
\includegraphics[scale = 0.52]{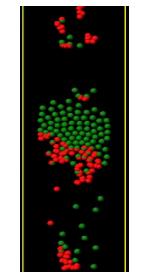} \\
$\rho~=~1\,$p/m$^2$ & $\rho~=~2\,$p/m$^2$ \\
  \hline
\end{tabular}
\caption{(Color on-line only) Snapshots of different evacuation processes for 
two desired velocities and two densities at the time-stamp of 15~s. Moving 
and fallen pedestrians are represented in green and red circles, respectively. 
The moving direction is from top to bottom. The yellow lines represent the 
walls of the pathway. The corridor width is $5\,$m.}
\label{fig:types_evacuation}
\end{figure}
\end{center}

Fig.~\ref{fig:types_evacuation} exhibits a few snapshots of the 
corridor simulations for different crowd densities. The moving 
direction is from from top to bottom. The corridor width is similar to the one 
in San Fermín (see caption for details). The overlapping individuals are 
actually the ones passing through fallen pedestrians.  \\

The first two columns in Fig.~\ref{fig:types_evacuation} correspond to the 
lowest value of the explored anxiety levels  ($v_d=3\,$m/s). We can see that no 
fallings occur for $\rho=~1~\,$p/m$^2$, but a unique cluster of fallen 
pedestrians 
appears for $\rho~=~2\,$p/m$^2$. An inspection of the animations shows that the 
first fallen runner triggers the falling of neighboring runners. This is in 
agreement with the falling events reported in the videos, and confirms once 
more, the relevancy of the first falling.\\

Interesting, there is an increase in the number of \textit{isolated} 
(\textit{i.e} not in contact) human clusters of fallen pedestrians for the 
more stressing scenario (last two columns of Fig.~\ref{fig:types_evacuation}). 
This is also in agreement with the empirical patterns exhibited in 
Fig.~\ref{tab:snapshots_real_video}. Furthermore, the complete video recordings 
show that stressed pedestrians tend to rush chaotically, and thus, the 
probability of fallings increases (see Fig.~\ref{fig:plot_proba}). \\

The fallings shown in Fig.~\ref{fig:types_evacuation} appear somehow related to 
the overall density of the crowd and the degree of clusterization during the 
run. Both issues will be studied separately in the next sections.

\subsubsection{Number of fallen pedestrians versus the density}
\label{sec:fallen_density}

\begin{figure}
\centering
\includegraphics[scale=1.0]{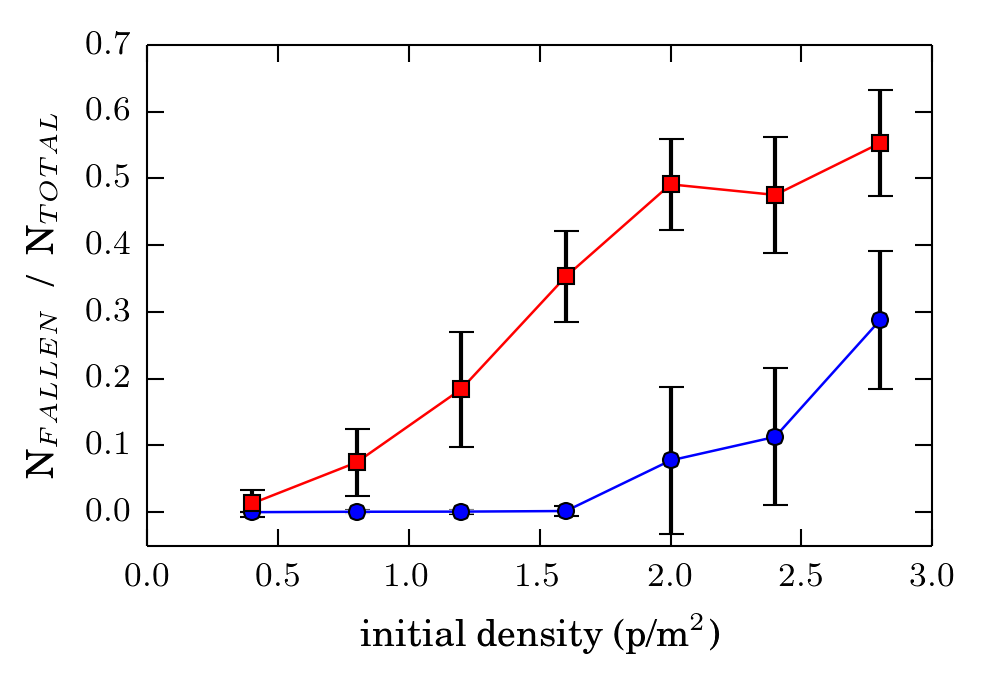}
\caption{\label{fig:fallen_density}(Color on-line only) Normalized number of 
fallen pedestrians as a function of the initial density for $v_d=3\,$m/s (blue 
circles) and $v_d=5\,$m/s (red squares) during the first 35~s (same corridor as 
in Fig.~\ref{fig:types_evacuation}). Initially, pedestrians were randomly 
distributed along the simulated corridor. Mean values were computed along 50 
realizations. The plot is normalized with respect to the total number of 
simulated pedestrians. The error bars corresponds to $\pm\sigma$ (one standard 
deviation).}
\end{figure}

Fig~\ref{fig:fallen_density} shows the number of stumbles 
occurring during the first 35~s of the escaping process as a function of 
the global density (see caption for details). We analyzed two desired 
velocities ($v_d=3\,$m/s and $v_d=5\,$m/s) since these appear as quite 
reasonable values according to data in \ref{appendix_1}.  \\

We can see in Fig.~\ref{fig:fallen_density} that the number of fallen 
pedestrians increases monotonically with density. This is an expected result 
since the more crowded environment, the more strikes between runners, and 
consequently, the more chances to fall down. These chances also increase with 
the desired velocity, that is, with the anxiety level of the individuals. Notice 
that these results are in agreement with \textit{in situ} videos (see 
Section~\ref{sec:experimental}). Specifically, the snapshots in 
Fig.~\ref{tab:snapshots_real_video} show that the stumbles commonly occur in 
the high density scenario (\textit{i.e.} in the post-corridor scenario). \\

We should call the attention on the fact that Fig.~\ref{fig:fallen_density} 
computes the fallings regardless of the clusterization context. This opens the 
question on whether the clusters formation tend to increase the number of 
fallings or not. We turn to this point in the following Section.

\subsubsection{Clusters}
\label{sec:size}

\begin{figure}
\centering
\includegraphics[scale=1.0]{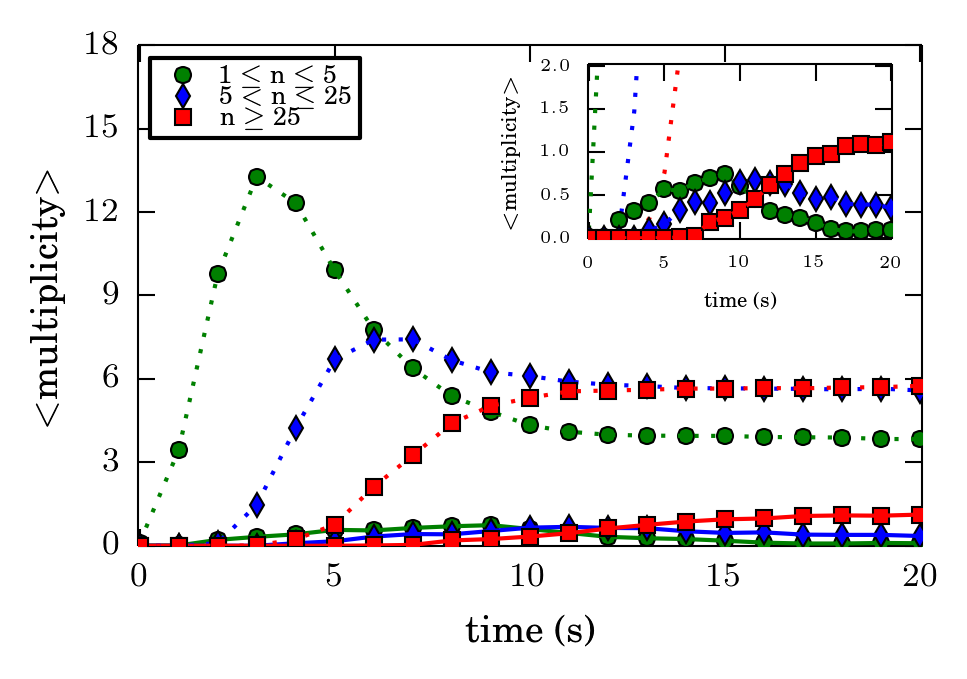}
\caption{\label{fig:plot_multiplicity}(Color on-line only) Mean multiplicity of 
human clusters of fallen pedestrians ($n$) for three different size 
categories (see legend for details) as a function of time. The size of each 
human cluster was obtained every $1\,$s. Data was acquired from 100 
simulation processes. Solid and dashed curves correspond to $v_d=3\,$m/s and 
$v_d=5\,$m/s, respectively. The density was $\rho~=~2\,$p/m$^2$.}
\end{figure}

We now turn to study the clustering structures, as defined in 
Section~\ref{human}. We classified the human clusters (of fallen 
pedestrians) into three categories (according to their size $n$) as 
follows: small ($1\leq n\leq5$), medium ($5<n<25$) and big ($n\geq25$). 
Fig.~\ref{fig:plot_multiplicity} shows the mean multiplicity of human clusters 
for each category (see caption for details). \\

A very first inspection of Fig.~\ref{fig:plot_multiplicity} shows that small 
human clusters have a major role along the whole process (for $v_d=5$), while 
its role at the early stage of the processes appears as essential. \\ 

A closer inspection of Fig.~\ref{fig:plot_multiplicity} shows the following:

\begin{itemize}
 \item Small human clusters ($1\leq n\leq5$): The occurrence of small human 
clusters is greater in the higher stress scenario ($v_d=5\,$m/s) than 
in the lower one ($v_d=3\,$m/s), as depicted previously in 
Fig.~\ref{fig:types_evacuation}. This means that, the higher the anxiety level, 
the most likely that (non contacting) pedestrians can initiate a falling 
process. This is a very dangerous environmental condition since any pedestrian 
becomes a potential instance of falling (for himself/herself and others). \\

\item Medium human clusters ($5<n<25$): The higher the anxiety level, the 
more likely to occur. This is somehow engaged to the existence of small clusters 
since the latter may increase its size (due to new fallings) as the process 
evolves. In other words, an examination of the process animations shows that 
small clusters can merge into a medium cluster.\\

\item Big human clusters ($n \geq 25$): A single big cluster can be observed in 
the low stress scenario, while six of them occur in the high stress one. As in 
the medium human clusters, an inspection of the animations shows that many 
medium clusters can merge into  bigger one. \\
\end{itemize}

We resume the highly stressing scenario as the one where fallings occur in many 
places and fallen clusters of any size are likely to occur. The low stressing 
scenario is substantially different since people fall around a single cluster. 
\\  

As a concluding remark, the massive stumble in the post-corridor event shown in 
Fig.~\ref{tab:snapshots_real_video} is similar to the above low stressing 
scenario, attaining a big fallen cluster. We presume that the runners at this 
instance were quite tired, resembling a low anxiety behavior (in the context of 
the SFM).

\subsubsection{Size of the largest avalanche}
\label{sec:largest}

\begin{center}
\begin{figure*}[!ht]
\hspace{0mm}
\subfloat[$\rho~=~2\,$p/m$^2$\label{fig:plot_velocity_N_300}]{\includegraphics[
width=0.5\columnwidth]{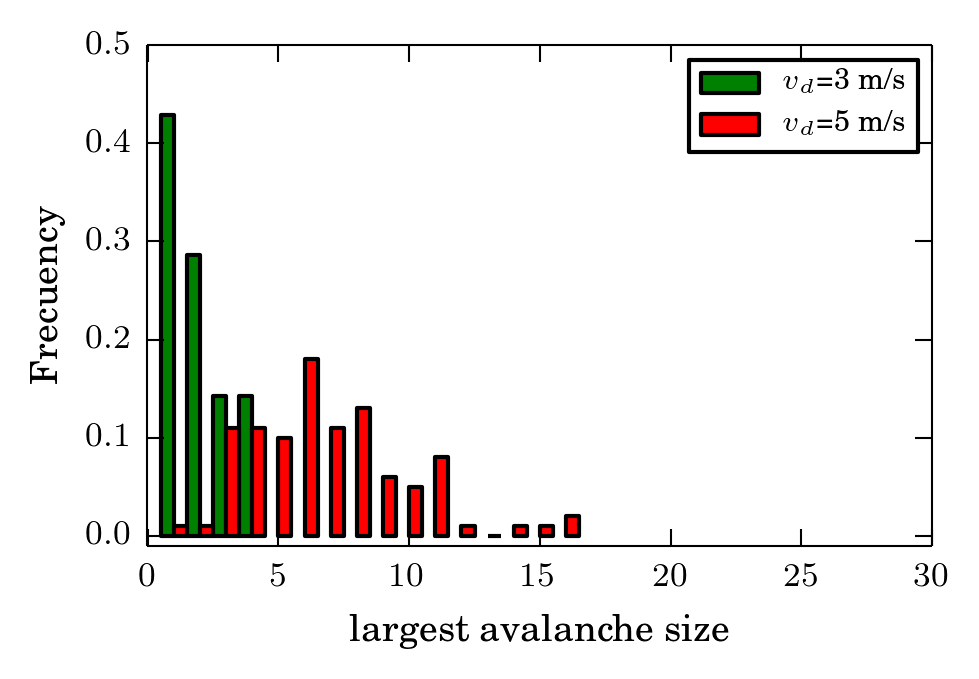}
}
\hspace{-1mm}
\subfloat[$\rho~=~3\,$p/m$^2$\label{fig:plot_velocity_N_700}]{\includegraphics[
width=0.5\columnwidth]{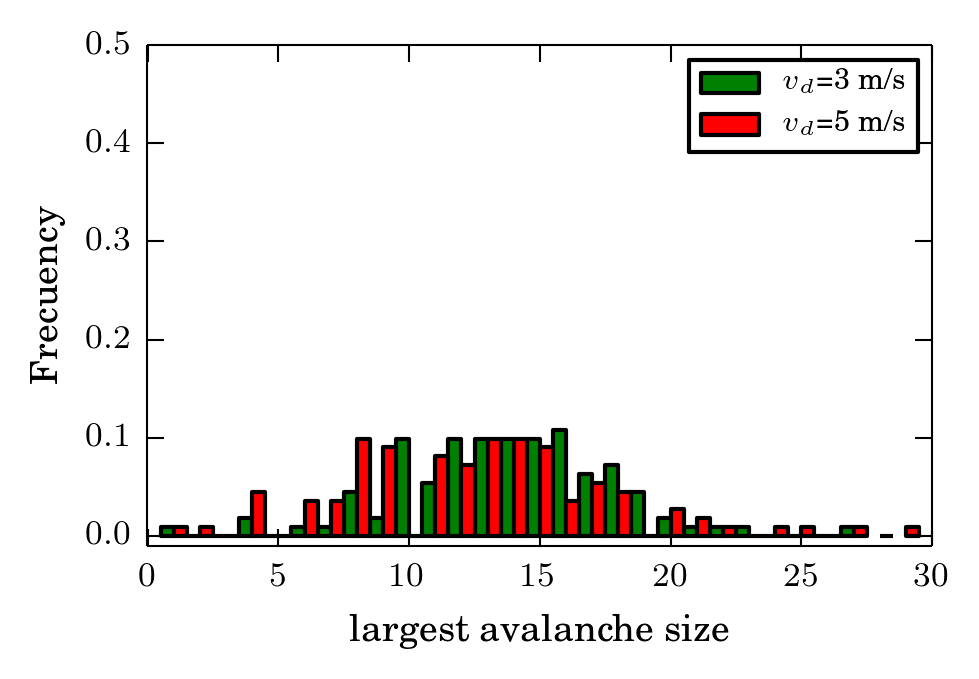}
}
\caption{\label{tab:cluster_vel} (Color on-line only) Normalized distribution 
of the largest avalanche size for two stress levels and two 
densities (see legend for details). The bin size is 5 pedestrians. The 
plot is normalized with respect to the total number of events. Data was 
recorded from 100 escaping processes. } 
\end{figure*}
\end{center}

Besides the size of the human clusters, we further examined the size of the 
largest avalanche of fallen pedestrians. That is, the maximum number of new 
fallen pedestrians (regardless of the falling location). This is, from our point 
of view, a meaningful magnitude to express the cluster formation rate. 
Fig.~\ref{tab:cluster_vel} shows the histograms of the largest avalanche 
distribution for two density levels (see caption for details). We omitted the 
case $\rho~=~1\,$p/m$^2$ because it does not provide enough falling samples for 
a proper statistic. We considered in Fig.~\ref{tab:cluster_vel} the densities 
$\rho~=~2\,$p/m$^2$ and $\rho~=~3\,$p/m$^2$. \\
 
We first notice that the mean value of the distribution depends on the 
pedestrian's anxiety level in the low-density situation ($\rho~=~2\,$p/m$^2$). 
As can be seen in Fig.~\ref{fig:plot_velocity_N_300}, the (maximum) avalanche 
events do not involve more than 5 pedestrians for $v_d=3\,$m/s, yielding a 
narrow distribution. But, in the highly anxious situation of $v_d=5\,$m/s, the 
corresponding distribution widens to allow up to 15 or 17 new fallings. This 
phenomenon appears reasonable since anxious pedestrians rush up to $v\approx 
v_d$, and thus, they become more susceptible to fallings ($f_s\approx 
v_d\,\nabla\rho$).\\

Besides, Fig.~\ref{fig:plot_velocity_N_700} shows that the avalanche events, 
mainly, involve 10 to 15 pedestrians for $\rho=3\,$p/m$^2$, regardless of the 
anxiety level. These are massive events that occur in a quite crowded 
environment, somewhat similar to the one shown in 
Fig.~\ref{tab:snapshots_real_video} for the post-corridor scene. As already 
observed in Section~\ref{sec:emp_meas}, these fallings occur because pedestrians 
embody striking situations (\textit{i.e.} high density gradients), regardless of 
the rushing level. Recall that high density gradients yield highly susceptible 
conditions for the falling. \\

We stress the fact that varying the anxiety level does not significantly 
change the behavior of the largest avalanche distribution. Or, in other 
words, the avalanches are mainly controlled by density regardless the anxiety 
level in high-density situation. 

\subsubsection*{\label{sec:rem_num}Concluding remarks from the numerical 
results}

The results from the numerical simulations show that highly dense environments 
tend to increase the number of fallings. The final number of fallings, however, 
is a complex interplay between the crowd density and the pedestrians anxiety 
level:   

\begin{itemize}
 \item For a moderate environmental density and low anxiety levels 
($\rho~=~2\,$p/m$^2$ and $v_d=3\,$m/s, respectively), the fallings are expected 
to occur in small avalanches at a single location. Thus, the picture is of a 
single cluster of fallen pedestrians. 

\item For a moderate environmental density but high anxiety levels 
($\rho~=~2\,$p/m$^2$ and $v_d=5\,$m/s, respectively), the fallings occur at many 
locations and can involve a varying number of falling people each.  

\item For more crowded environments (say, $\rho~=~3\,$p/m$^2$), the anxiety 
level looses significance since the pedestrians strike among others, no matter 
their escaping desire.
\end{itemize}


\section{Conclusions}\label{sec:conclusions}

Our research focused on the microscopic analysis of the stumbling phenomena 
during an emergency situation (the Running of th Bulls' Festival at San Ferm\'\i 
n). We first analyzed three real life videos of the rush and proposed a 
stumbling mechanism in the context of the SFM. We then performed \textit{ex 
post} simulations for evaluating the accuracy of the model improvements. \\

In order to properly understand the stumbling mechanism, we examined the crowd 
behavior before and after the first falling event. This allowed us to 
distinguish between an homogeneous scenario (with ``moving'' 
pedestrians) and an heterogeneous one (with ``moving'' and ``fallen'' 
pedestrians). \\

We identified the product between the local density gradient and the 
velocity of each pedestrian (named here as susceptibility $f_s$) as a good 
indicator for an imminent fall. We noticed that $f_s$ is monotonically related 
to the falling probability. This relation can be found in 
Section~\ref{sec:emp_meas}. \\

We further noticed through the experimental data that the presence of fallen 
pedestrians dramatically increases the falling probability of neighboring 
pedestrians. Our computations show that the latter have, at least, an 
additional $20\%$ of chances for falling down. Thus, the presence or not of a 
fallen neighbor is a crucial feature of pedestrians (local) environment. This is 
actually the main achievement of our investigation.\\

The pedestrian's own susceptibility and the presence (or not) of fallen 
neighbors are the only features allowing a stumble within the context of the 
Social Force Model. Although the simplicity of this model, we attained fairly 
good agreement with empirical data from the San Ferm\'\i n Festival.   \\

We performed \textit{ex-post} simulations of the ``Running of the 
Bulls''. We used different sets of densities ($\rho$) and anxiety levels 
($v_d$) for mimicking three instances of the race (see 
Section~\ref{sec:num_sim}). Our results reproduced satisfactorily the 
corresponding video recordings for adequate values of $\rho$ and $v_d$. We 
further focused on $v_d=3\,$m/s and $v_d=5\,$m/s for a detailed analysis of the 
stumbling process (see Section~\ref{sec:num_sim}). The conclusions are as 
follows:

\begin{itemize}
 \item The moderately dense scenarios ($\rho~=~2\,$p/m$^2$) may evolve quite 
differently, according to the anxiety level. If $v_d=3\,$m/s (low anxiety 
level) a small stumbling process can takes place at a single location.  In this 
case, we will observe  single cluster of fallen pedestrians. However, if 
$v_d=5\,$m/s (high anxiety level) the stumbling process may take place in 
multiple places within the crowd. Indeed, we observe a varying number of people 
at falling and clusters of fallen pedestrians among the crowd.

\item The highly dense scenario ($\rho~=~3\,$p/m$^2$) may evolve in the same 
way, no matter the anxiety level (within our explored range). This way will show 
up as a massive avalanche within the crowd. 
\end{itemize}

\section*{Acknowledgments}
C.O.~Dorso is a main researcher of the National Scientific and Technical 
Research Council (spanish: Consejo Nacional de Investigaciones Cient\'\i ficas 
y T\'ecnicas - CONICET), Argentina and full professor at Departamento de 
F\'isica, Universidad de Buenos Aires. G.A.~Frank is a researcher of 
the CONICET, Argentina. F.E.~Cornes is a PhD Student in Physics. This work was 
supported by the Fondo para la investigaci\'on cient\'ifica y tecnol\'ogica 
(FONCYT) grant Proyecto de investigaci\'on cient\'ifica y tecnol\'ogica Number 
PICT-2019-2019-01994. G.A.~Frank thanks Universidad Tecnológica Nacional (UTN) 
for partial support through Grant PID SIUTNBA0006595.  \\


\appendix

\section{\label{appendix_1}Velocity and local density gradient histograms}

\begin{center}
\begin{figure}[!ht]
\begin{tabular}{|@{\hspace{0mm}}c|@{\hspace{0mm}}c|@{\hspace{0mm}}c|}
\multicolumn{3}{c}{Velocity}\\
\hline
\includegraphics[scale = 0.35]{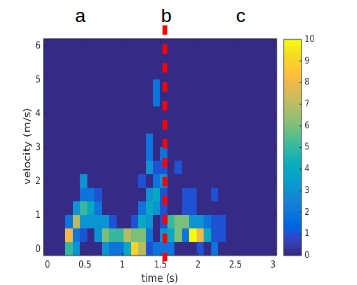} & 
\includegraphics[scale = 0.23]{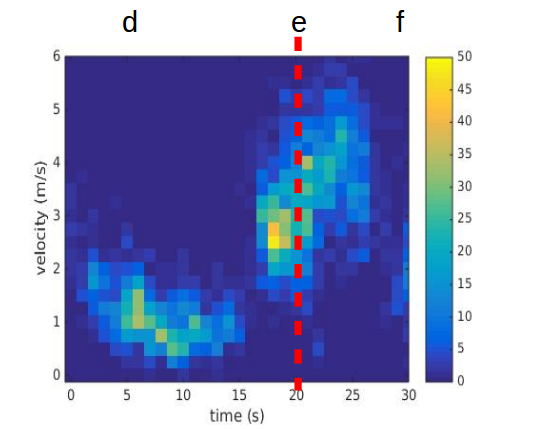} & 
\includegraphics[scale = 0.23]{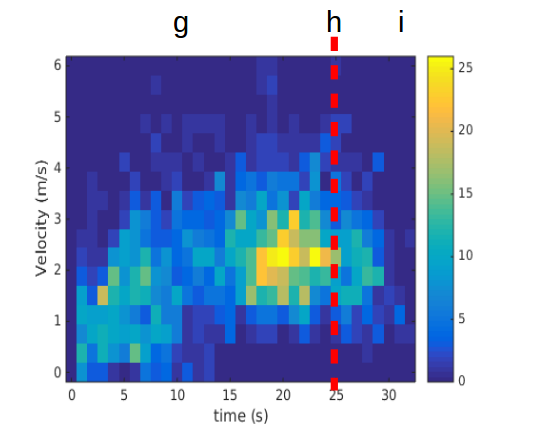} \\
Pre-corridor & Corridor & Post Corridor \\
  \hline
  \multicolumn{3}{c}{Local density gradient}\\
  \hline
\includegraphics[scale = 0.22]{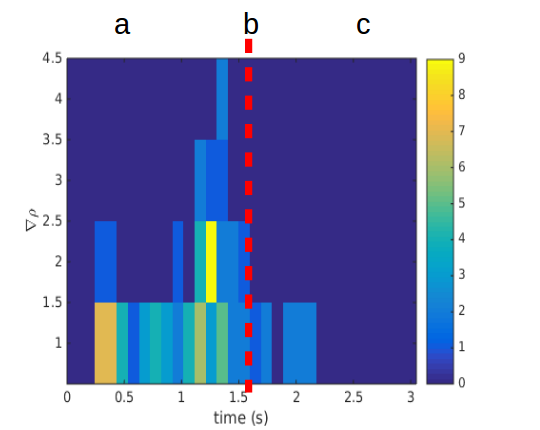} & 
\includegraphics[scale = 0.22]{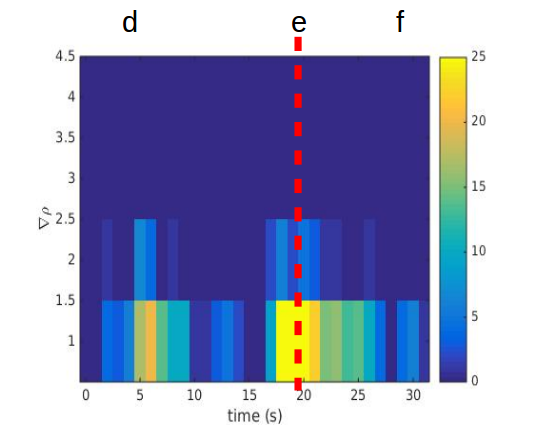} & 
\includegraphics[scale = 0.22]{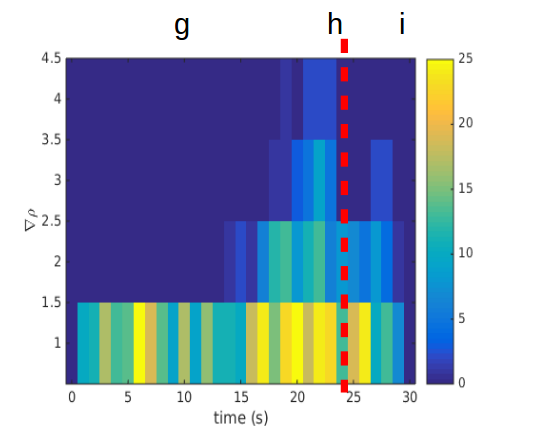}\\
Pre-corridor & Corridor & Post Corridor \\
  \hline
\end{tabular}
\caption{(Color on-line only) 2-D histograms for the velocity (upper) and local 
density gradient (lower) for each scenario. The horizontal and vertical axis 
represent the time and the measured value of each magnitude, respectively. The 
bin size was $0.125\,\mathrm{s}\times0.33\,\mathrm{m/s}$ and 
$0.125\,\mathrm{s}\times1\,\mathrm{people}$ for the velocity and the local 
density gradient, respectively. The vertical red dashed line represents the 
time of the first fall. Also, for clarity reasons, the upper limit of each 
color bar is different. The crowd states at different moments (see letters 
above figures) can be seen in Fig.~\ref{tab:snapshots_real_video}.}
\label{tab:histogramas}
\end{figure}
\end{center}

Fig.~\ref{tab:histogramas} shows the velocity histograms as a function of time 
for each scenario (see  caption for details). We observe that, likewise the 
local density pattern (see 
Figs.~\ref{fig:densidad_pre_corridor}-\ref{fig:densidad_entrada}), the 
three scenarios present a common behavior: the first fall occurs \emph{after} 
an increase in the number of pedestrians attaining high velocity (say, 
$v>2\,$m/s). 
\\

A careful inspection of the scenes \emph{before} the falls shows that 
some pedestrians moving with high velocity trigger the run of many other 
relaxed individuals. Consequently, this ``fear spreading'' among the crowd 
makes the maximum of the velocity distribution to move towards higher values.\\

The local density gradient histograms along time for each scenario can be found 
in Fig.~\ref{tab:histogramas} (see caption for details). Notice that 
this magnitude carries information on the local density profile around each 
pedestrian (see Eq.~\ref{eqn_grad}). That is, the higher the local density 
gradient, the greater the local density asymmetry in the front–back direction. 
And therefore, the higher the force unbalance on the corresponding 
pedestrian.\\

It can be noticed that there is an increment in the occurrence of high local 
density gradients (say, $\nabla\rho>1$) few seconds \emph{before} 
the first falling. In this sense, in the pre-corridor and post-corridor 
scenarios, a sizable portion of the pedestrians experienced a significant amount 
of unbalance in the front-back forces (say, $\nabla\rho\geq3$). Notice that this 
is not significant in the corridor scenario. 

\section*{References}

\bibliographystyle{elsarticle-num}
\bibliography{paper}

\end{document}